\documentclass[sigconf,natbib=false]{acmart}
\AtBeginDocument{%
  \providecommand\BibTeX{{%
    Bib\TeX}}}

\setcopyright{acmlicensed}
\copyrightyear{2025}
\acmYear{2025}
\acmDOI{XXXXXXX.XXXXXXX}

\acmConference[MM'25]{33rd ACM International Conference on Multimedia}{October 27--31, 2025}{Dublin, Ireland}
\acmISBN{978-1-4503-XXXX-X/2025/10}

\RequirePackage[
  datamodel=acmdatamodel,
  style=acmnumeric,
  ]{biblatex}

\begin{document}

\title{KLASSify to Verify: Audio-Visual Deepfake Detection Using SSL-based Audio and Handcrafted Visual Features}

\author{Ivan Kukanov}
\authornote{Both authors contributed equally to this research.}
\email{ivan@kukanov.com}
\orcid{1234-5678-9012}
\author{Jun Wah Ng}
\authornotemark[1]
\email{junwah.ng@klasses.com.sg}
\affiliation{%
  \institution{KLASS Engineering and Solutions}
  \country{Singapore}
}

\newcommand{\red}[1]{\textcolor{red}{#1}}
\newcommand{\ivan}[1]{\textcolor{purple}{\textbf{IVAN:} #1}}
\newcommand{\todo}[1]{\textcolor{cyan}{\textbf{TODO:} #1}}
\newcommand{\lit}[1]{\textcolor{red}{[\textbf{ref:}#1]}}

\begin{abstract}
The rapid development of audio-driven talking head generators and advanced Text-To-Speech (TTS) models has led to more sophisticated temporal deepfakes. These advances highlight the need for robust methods capable of detecting and localizing deepfakes, even under novel, unseen attack scenarios. Current state-of-the-art deepfake detectors, while accurate, are often computationally expensive and struggle to generalize to novel manipulation techniques. To address these challenges, we propose multimodal approaches for the AV-Deepfake1M 2025 challenge. For the visual modality, we leverage handcrafted features to improve interpretability and adaptability. For the audio modality, we adapt a self-supervised learning (SSL) backbone coupled with graph attention networks to capture rich audio representations, improving detection robustness. Our approach strikes a balance between performance and real-world deployment, focusing on resilience and potential interpretability. On the AV-Deepfake1M++ dataset, our multimodal system achieves AUC of $92.78\%$ for deepfake classification task and IoU of $0.3536$ for temporal localization using only the audio modality. 
\end{abstract}



\begin{teaserfigure}
\centering
  \includegraphics[scale=0.3]{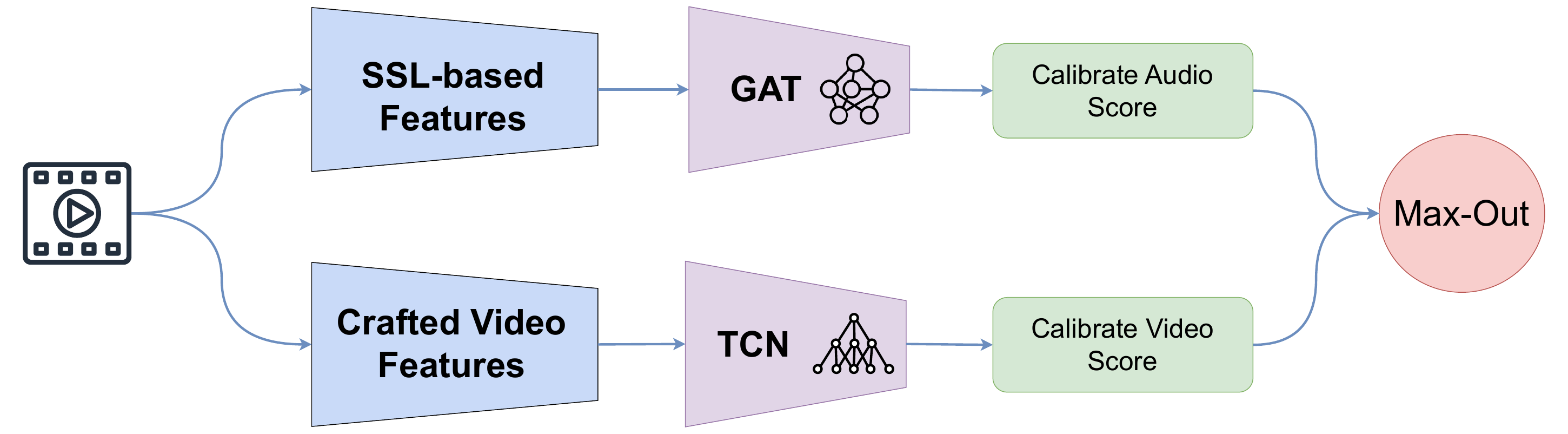}
  \caption{KLASS Solution Pipeline for Deepfake Classification. Self-supervised Learning (SSL)-based features followed by graph attention networks (GAT) are fine-tuned for audio stream, crafted video features are trained with temporal convolution networks (TCN). The prediction scores from audio-visual models are calibrated and \emph{max-out} decision is obtained per video clip.}
  \label{fig:teaser}
\end{teaserfigure}

\received{1 August 2025}

\maketitle

\section{Introduction}
The advancement of audio-driven talking head generators has made it trivial to create temporal deepfakes, where only the speech segment and the corresponding mouth-region motion are manipulated, while the remainder of the video stays authentic. State-of-the-art talking head generators \cite{ li2025latentsynctamingaudioconditionedlatent, mukhopadhyay2023diff2lipaudioconditioneddiffusion,wang2023seeingsaidtalkingface} can produce frame-level lip movements that are indistinguishable from real videos. More importantly, video deepfake manipulations are now accompanied by equally powerful TTS systems. Zero-shot TTS models \cite{casanova2023yourttszeroshotmultispeakertts,casanova2024xttsmassivelymultilingualzeroshot} can closely clone a speaker's voice without requiring extra fine-tuning. When sufficient data are available, TTS models such as VITS \cite{kim2021conditionalvariationalautoencoderadversarial} can be further fine-tuned to produce intonation and prosody that are nearly indistinguishable from the real speaker. These new waves of deepfake attacks are poised to amplify disinformation, erode trust in audiovisual evidence, and threaten national security.

Large-scale temporal deepfake benchmark datasets \cite{cai2024av,CAI2023103818, cai2025avdeepfake1mlargescaleaudiovisualdeepfake} highlight both the scale and the challenges faced by existing methods when detecting short manipulated segments embedded in long, genuine videos. The deepfake research community has increasingly focused on multi-modal temporal detectors that jointly analyze audio and video streams. Existing state-of-the-art multimodal temporal deepfake detectors \cite{BATFD+, BATFD, Zhang_2023_UMMAFormer} typically rely on large, pre-trained encoders and complex fusion layers to detect and localize deepfakes. These heavy backbones, often with hundreds of billions of parameters, achieve state-of-the-art accuracy but require significant computational resources, making them impractical for real-time deployment. However, such models often suffer from poor robustness to distribution shifts, unseen or new manipulations, of which are common in real-world forensics applications. An example would be BA-TFD+ \cite{BATFD+}, which scored 96.30 AP@0.5 on the LAV-DF \cite{CAI2023103818} but struggles at 14.7 AP@0.5 in the AV-Deepfake1M++ \cite{cai2025avdeepfake1mlargescaleaudiovisualdeepfake} testA split. 
To address these challenges, we propose multimodal approaches for both the detection and localization tracks of the AV-Deepfake1M 2025~\cite{cai2025avdeepfake1mlargescaleaudiovisualdeepfake} competition. We first decouple the audio and video modalities, allowing independent analysis and providing more interpretable insights. For the visual modality, we extract handcrafted features that capture the visual artifacts inherent in deepfakes. These features enhance both explainability and robustness to unseen attacks. For the audio modality, we leverage SSL-based features, enabling us to capture complex, high-level audio cues such as intonation, prosody, and phonetic patterns. 
These features are particularly effective in detecting manipulated audio in deepfakes \cite{Tak2022Automatic}, even when faced with unseen types of attacks. This improves the model's resilience to new and evolving deepfake manipulation techniques, ensuring reliable detection across diverse audio manipulations.

\textbf{Detection Task}
For Task 1, we propose a simple score-based shallow fusion approach that achieved high accuracy in the detection task. By training a TCN \cite{lea2016temporalconvolutionalnetworksunified} on handcrafted visual features alongside Wav2Vec-AASIST \cite{Tak2022Automatic} trained on audio features, we were able to achieve competitive results.

\textbf{Localization Task}  For Task 2, the boundary-aware attention \cite{zhong2024enhancingpartiallyspoofedaudio} approach is adapted for localizing fake segments in the audio stream.

\section{Problem Statement}
\textbf{Tasks Overview}. The proposed methods were developed according to the AV-Deepfake1M 2025~\cite{cai2025avdeepfake1mlargescaleaudiovisualdeepfake} challenge rules. The challenge comprises two tasks supported by the AV-Deepfake1M++ \cite{cai2025avdeepfake1mlargescaleaudiovisualdeepfake} dataset:

\textbullet{} Task 1. Given a video clip featuring a single speaker, the objective is to determine whether the video is real or manipulated. For this task, participants are restricted to using video-level labels for training. 

\textbullet{} Task 2. Given similar video samples, the goal is to identify time stamps of manipulated fake segments induced in real video. Participants are provided with the ground-truth segment-level labels that indicate the exact locations of the fake content.


\textbf{Evaluation Metrics}.
For Task 1, the evaluation metric for deepfake detection is the Area Under the Curve (AUC) score. The AUC measures the discrimination ability of a deepfake detection system on multiple operating points. For Task 2, the metrics are the Average Precision (AP) and the Average Recall (AR). The overall score is computed as:
\begin{equation}
\begin{aligned}
    \text{Score} &= \frac{1}{8} \sum_{\text{IoU} \in \{0.5, 0.75, 0.9, 0.95\}} \text{AP@IoU} \,\, + \\
    &\quad \frac{1}{10} \sum_{\text{N} \in \{50, 30, 20, 10, 5\}} \text{AR@N}
\end{aligned}
\end{equation}

\section{Related Work}
\textbf{Visual deepfake}
Early visual deepfake detectors operated frame-by-frame, focusing on spatial artifacts within individual frames \cite{zheng2021exploringtemporalcoherencegeneral}. However, since most deepfakes are generated in a frame-by-frame manner, they often exhibit clear temporal inconsistencies that single-frame detectors fail to capture. To address this, recent approaches have incorporated temporal modeling to detect inter-frame anomalies such as flickering, jitter, or abrupt shifts in facial landmarks \cite{gu2021spatiotemporalinconsistencylearningdeepfake}. Studies have shown that audio-driven lip-sync deepfakes frequently disrupt the smooth, natural kinematics of lip movement, with synthetic fakes failing to adhere to normal physiological dynamics, resulting in irregular motion patterns \cite{liu2025exposingtheforgeryclues}. Spatiotemporal detectors such as UMMAFormer \cite{Zhang_2023_UMMAFormer} leverage frame differences across both RGB and optical flow to capture such inconsistencies. Other approaches, like \cite{Tan2023deepfakefacialaction}, model facial action unit dynamics to detect irregularities, improving the detection of speech-driven and reenactment fakes.
Most detectors lack generalizability, performing well on known generators but struggling with unseen ones due to overfitting to generator specific artifacts \cite{xiong2025talkingheadbenchmultimodalbenchmark}. Despite recent advances, common artifacts persist across generators. Early face swaps often displayed boundary and lighting mismatches \cite{xiong2025talkingheadbenchmultimodalbenchmark}, while modern talking-head generators continue to struggle with regional consistency. Many lip-sync fakes alter only the mouth region, leaving the surrounding face unnaturally static during speech \cite{datta2024exposinglipsyncingdeepfakesmouth}. Additionally, forged mouth regions often exhibit blurriness, color shifts, or unnatural landmark kinematics.

These cross-generator artifacts motivate our use of handcrafted temporal features focused on the mouth region. Our approach is based on the intuition that even as deepfake algorithms evolve, subtle anomalies in temporal coherence persist. By targeting these universal artifacts, rather than generator specific cues, we aim to enhance robustness against unseen attacks.

\textbf{Speech deepfake classification}
Deepfake audio detection aims to differentiate between genuine human speech and speech that has been artificially generated with TTS system. Progress in this field has been significantly driven by the ASVspoof challenge series \cite{Todisco2019ASVspoof2F, Wang2024ASVspoof5C, ASVspoof2021}, which offer standardized benchmarks and evaluation protocols. Initial approaches often relied on manually crafted acoustic features, but the field soon shifted toward deep learning models, notably CNN architectures like Light CNN (LCNN) \cite{lavrentyeva2019stcantispoofingsystemsasvspoof2019}. Trained from scratch on specific data corpus, these systems do not guarantee generalization across domains and acoustic environments \cite{müller2024doesaudiodeepfakedetection}. Recent advances have focused on leveraging large self-supervised learning (SSL) models - such as \texttt{Wav2Vec-2.0} \cite{Baevski_2020}, \texttt{HuBERT} \cite{HsuHubert}, \texttt{WavLM} \cite{chen2021wavlm} - which are known for their ability to capture rich high-level speech representations. A representative system is Wav2Vec-AASIST \cite{Jung_2022,Tak2022Automatic}, which combines a \texttt{Wav2Vec-2.0} encoder with a graph-attention-based classification head. This model is specifically designed for deepfake audio detection and has demonstrated strong benchmark performance on ASVspoof 2019~\cite{Tak2022Automatic,Todisco2019ASVspoof2F} where the whole utterance either bona fide or fully spoofed. 

However, there is no research on the Wav2vec-AASIST in \textit{weakly supervised}~\cite{Zhu05} scenarios where labels only indicate the presence~/~absence of fake segments within an audio, without the exact timing of the fake segment. 
The underlying components of the Wav2vec-AASIST - SSL-based features and the graph attention mechanisms - are potentially extendable to weakly supervised settings.

\textbf{Speech deepfake localization}
The problem of localizing fake segments was initially addressed in a series of PartialSpoof research \cite{zhang23_partialspoof_taslp,zhang21_asvspoof,zhang21ca_interspeech}. Where the training is fully-supervised frame-based with real/fake labels. The task of partially spoofed audio localization aims to accurately determine audio authenticity at a frame level, determine time stamps and a probability of segment being fake. 
In \cite{zhang23_partialspoof_taslp}, the proposed approach employed self-supervised learning (SSL) models for audio feature extraction and a multi-resolution head (20ms to 640ms) architecture capable of detecting spoofing at both utterance and segment levels simultaneously. 
The approach with the \emph{boundary-aware attention mechanism} (BAM) \cite{zhong2024enhancingpartiallyspoofedaudio} showed strong performance on PartialSpoof \cite{zhang21_asvspoof} and PartialEdit \cite{zhang2025PartialEdit} datasets. We adapted the BAM approach for the Task 2 audio deepfake localization of this challenge.

\section{Methodology}

\subsection{Deepfake Classification}
\label{sec:deepfake-classification}

\textbf{Video Modality}
We empirically observed that deepfake temporal lip-sync attacks frequently exhibit distinctive visual artifacts.
For each video, we use the Face Mesh model from MediaPipe \cite{lugaresi2019mediapipe} to compute facial landmarks and regions of interest (ROIs), from which we extract handcrafted features. We identified eight key discriminative features, residual artifacts commonly left by lip-sync generators:
\begin{enumerate}
\item \textbf{Blurriness of the Mouth ROI} – Increased blurriness in the mouth region, measured by the variance of the Laplacian.
\item \textbf{Non-mouth Mean Squared Error (MSE)} – MSE between consecutive frames in the \textit{non-mouth} regions of the face. Since generators focus on lip movement, the rest of the face and background often remain unnaturally static \cite{wiles2018x2facenetworkcontrollingface}.
\item \textbf{Color Shift of the Mouth ROI} – Frame-to-frame color change in the mouth region, computed in the Lab color space. This captures discrepancies in skin tone between real and generated frames.
\item \textbf{Landmark Kinematics} – Lip-sync generators frequently violate the natural kinematics of human speech. We quantify this using:
    \begin{itemize}
    \item \textbf{Mouth Aspect Ratio} – Ratio of vertical mouth opening to its width.
    \item \textbf{Velocity} – Real speech exhibits natural rates of mouth motion, while fakes often show erratic or overly smooth movements.
    \item \textbf{Acceleration} – Highlights dynamic changes, such as smooth motion versus abrupt shifts.
    \item \textbf{Jerk} – Measures motion smoothness. High-frequency jerk is a common deepfake artifact caused by frame-by-frame synthesis \cite{zheng2021exploringtemporalcoherencegeneral}.
    \item \textbf{Jitter} – Deviation of landmark trajectories from a linear motion model.
    \end{itemize}
\end{enumerate}

\begin{figure}
    \centering
    \includegraphics[scale=0.7]{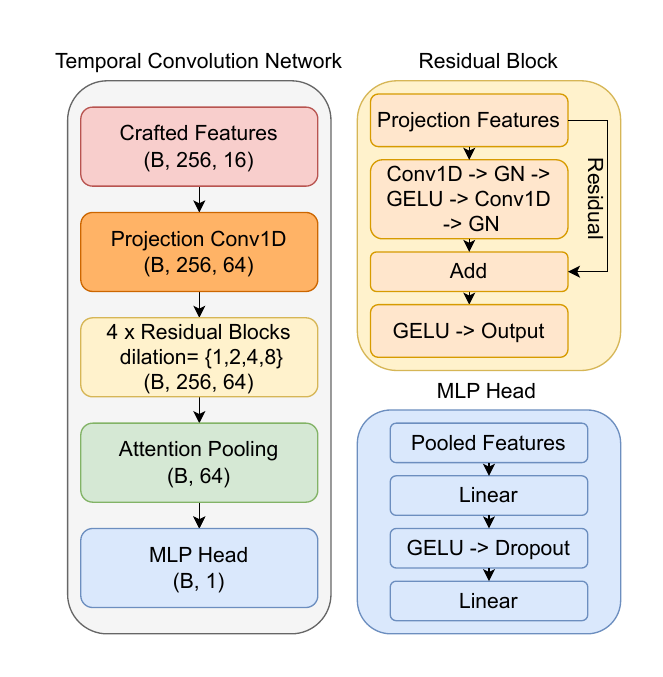}
    \caption{Architecture of the lightweight TCN~\cite{lea2016temporalconvolutionalnetworksunified} used for deepfake classification}
    \label{fig:tcnfigure} 
\end{figure}

Using these features, we train a lightweight 1D TCN~\cite{lea2016temporalconvolutionalnetworksunified} with only 124K parameters for video classification. In Figure~\ref{fig:tcnfigure}, the architecture projects the input features into a higher-dimensional space, processes them through four stacked residual blocks with increasing dilation factors to capture long-range temporal dependencies, and aggregates the output via attention pooling. A final multi-layer perceptron (MLP) performs binary classification.

\textbf{Audio Modality}
For the deepfake classification with partial fake segments, we adopt Wav2Vec-AASIST~\cite{Tak2022Automatic}, building upon prior work in speech deepfake detection \cite{Kukanov_2024,Tak2022Automatic}. Specifically, the architecture uses the Wav2Vec 2.0 XLSR-53 \cite{Baevski_2020} model (with a 1024-dimensional output) as the front-end feature extractor, followed by the AASIST, a spectro-temporal graph attention network, as the classifier. 

\textbf{Audiovisual Fusion}
The audiovisual fusion is performed at the score level, known as shallow fusion as shown in Figure \ref{fig:teaser}. The output of the video and audio models is softmax scores in the range of $[0, 1]$.  To avoid biased decisions by a certain modality, we train the probability calibration using Platt \cite{Platt1999Probabilistic} sigmoid scaling:
\begin{equation}
    p(y_i = 1 | s_i) = \frac{1}{1 + \text{exp}(as_i + b)},
\end{equation}
where $y_i$ is the true label of a sample $i$, and $s_i$ is the output score of the un-calibrated audio or video classifier for a sample $i$. The real-valued parameters $a$ and $b$ are trained separately for each modality. Finally, \emph{Max-Out} operation is applied to the calibrated audio $p_a$ and video $p_v$ probability scores to make a final video-level decision. 

\subsection{Deepfake Localization}

\textbf{Video Modality}
\begin{figure}
    \centering
    \includegraphics[scale=0.6]{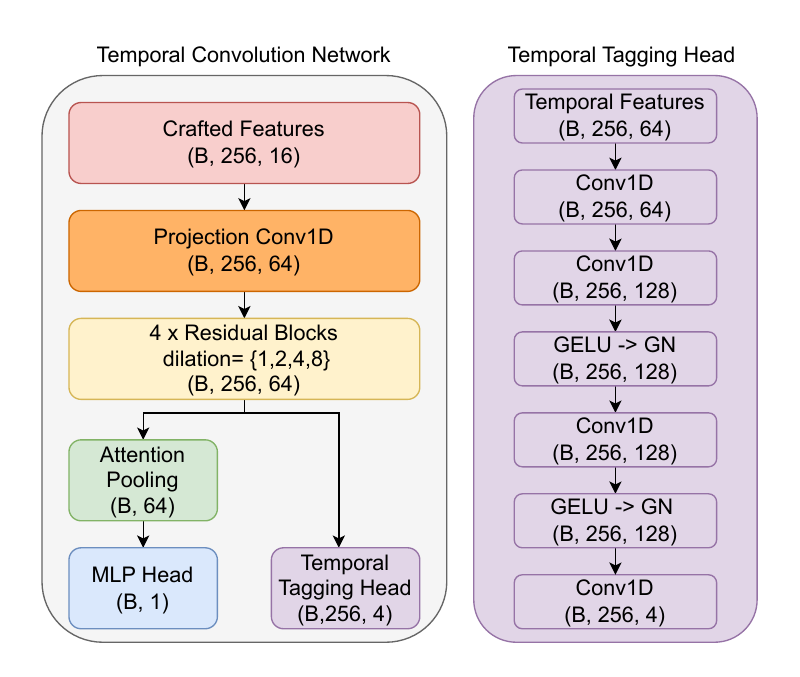}
    \caption{Architecture of the lightweight TCN~\cite{lea2016temporalconvolutionalnetworksunified} with a tagging head for temporal deepfake localization.}
    \label{fig:temporaltcnfigure} 
\end{figure}
The backbone explored for deepfake segments localization is identical to that used in classification. We design a lightweight yet expressive temporal tagging head that employs depthwise separable convolutions to efficiently model temporal dependencies, in Figure \ref{fig:temporaltcnfigure}. The addition of tagging head to TCN~\cite{lea2016temporalconvolutionalnetworksunified} increased the parameter count to 140K. The temporal decoder head produces logits for each frame across four tags - \textit{O (Outside)}, \textit{B (Begin)}, \textit{I (Inside)}, and \textit{L (Last)} - adopting the BILOU \cite{ratinov-roth-2009-design-bilou} tagging scheme to enhance the model's ability to identify the exact boundaries of the manipulated segments. 


\textbf{Audio Modality}
In this work, we adapt the \emph{Boundary-aware Attention Mechanism} (BAM) \cite{zhong2024enhancingpartiallyspoofedaudio} which shows the state of the art performance on the PartialSpoof~\cite{zhang21ca_interspeech} dataset. The approach comprises pre-trained SSL front-end feature extraction from WavLM~\cite{journals/corr/abs-2110-13900}, followed by two modules \emph{boundary enhancement} (BE) and \emph{boundary frame-wise attention} (BFA). The BE addresses the instability in detecting transition boundary frames in partial deepfakes. In such cases, frames containing a mix of real and fake content are labeled as fake. However, frames dominated by real content may closely resemble genuine frames in the feature space, leading to training instability. The \emph{boundary frame-wise attention} module leverages the boundary predictions to make more reliable frame-level decisions within and outside fake segments. Finally, the outputs from the BE and BFA modules are combined to produce the final frame-level authenticity predictions.

\section{Experimental Setup}

\subsection{Dataset}

\begin{figure}[!htb]
  \centering
  \includegraphics[width=\linewidth]{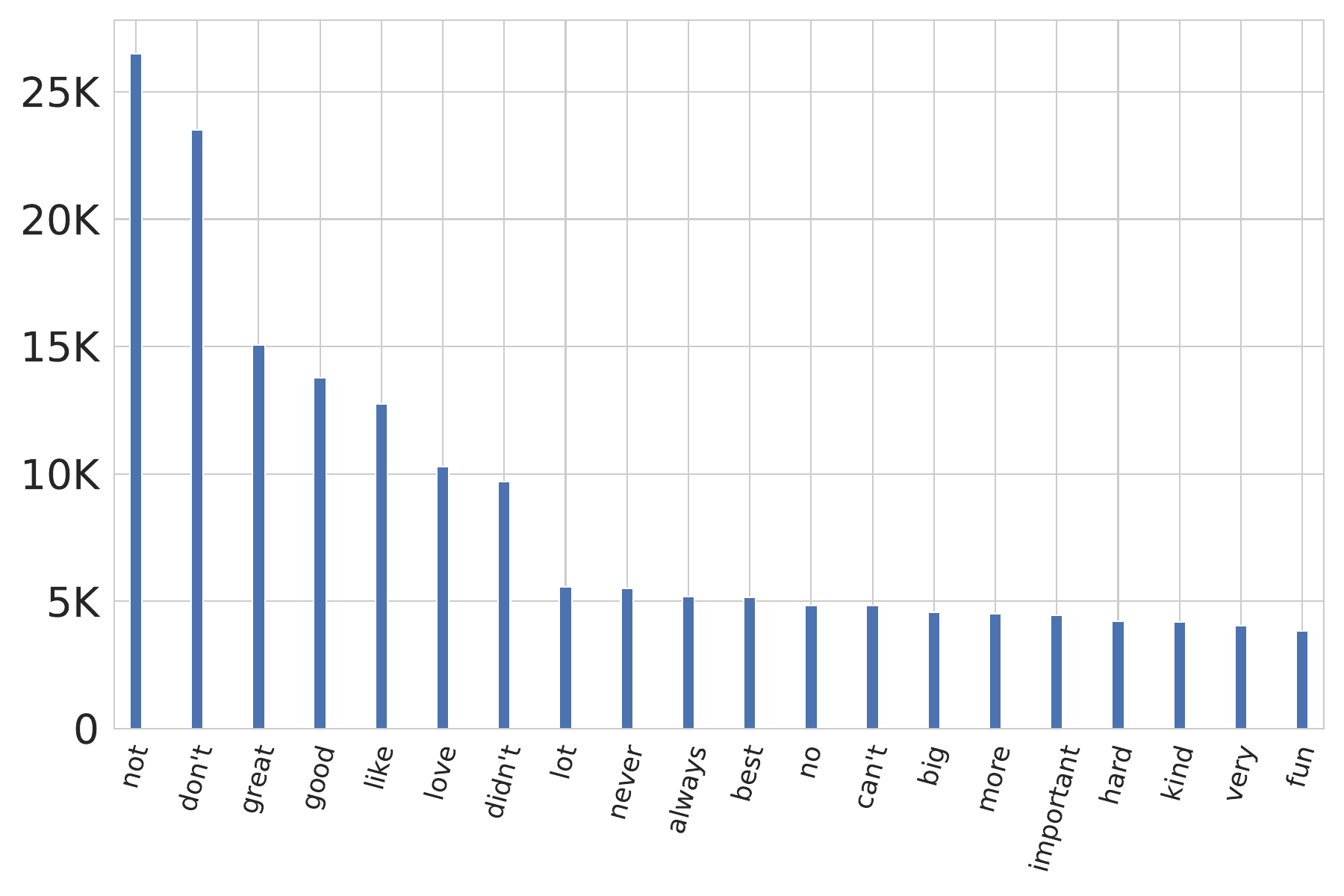}
  \caption{Histogram of the most frequently edited words in a sentence}
  \label{img:hist_word_edit}
\end{figure}

The total AV-Deepfake1M++ \cite{cai2025avdeepfake1mlargescaleaudiovisualdeepfake} dataset contains more than 2M videos. The training dataset consists of 1.1M videos, the validation set contains 77K videos, see Table \ref{tab:trainval-analysis}. Analyzing the dataset, it was noticed that most of the videos in the validation set are perturbed versions of videos already present in the training set. The training and validation sets are evenly distributed across four labels of \textit{real}, \textit{audio\_modified}, \textit{visual\_modified} and \textit{both\_modified}. The number of fake segments ranges from 0 to 5, with the mean duration of 0.33s, equivalent to a short word. The meaning of sentences was manipulated by editing context-sensitive words; Figure \ref{img:hist_word_edit} shows the most frequently modified words. More details on the \textit{testA} and \textit{testB} evaluation splits of the AV-Deepfake1M++ dataset can be found in \cite{cai2025avdeepfake1mlargescaleaudiovisualdeepfake}.
\begin{table}
  \caption{Training and Validation Dataset Statistics}
  \label{tab:trainval-analysis}
  \begin{tabular}{lcc}
    \toprule
    Statistics & Train & Val\\
    \midrule
    \emph{Data Distribution}\\  
    total & 1,099,217 & 77,326 \\
    real & 297,389 (27.05\%) & 20,220 (26.15\%) \\ 
    audio\_modified & 266,116 (24.21\%)  & 18,938 (24.49\%) \\ 
    visual\_modified & 269,901 (24,55\%) & 19,099 (24.70\%) \\ 
    both\_modified & 265,812 (24.18\%) & 19,069 (24.66\%) \\
    \midrule
    \emph{Duration of Videos} \\  
    minimum  & 1.32s & 2.52s \\
    maximum  & 668.48s & 152.40s \\
    mean  & 9.60s & 9.56s \\
    \midrule
    \emph{Number of Attacks per Video} \\
    Range & 0 to 5 & 0 to 5 \\
    \midrule
    \emph{Duration of Fake Segments} \\
    minimum  & 0.02s & 0.02s \\
    maximum  & 11.58s & 8.10s \\
    mean & 0.33s & 0.33s \\
    \midrule
  \bottomrule
\end{tabular}
\end{table}

\begin{table}[!htb]
  \caption{Deepfake Classification Results (Task 1) on the validation split (\emph{Val}) and \emph{TestA} of AV-Deepfake1M++ \cite{cai2025avdeepfake1mlargescaleaudiovisualdeepfake} dataset. The validation results for the \textit{Audio} and \textit{Video} models are presented on their respective modality scores only}
  \label{tab:classif-result}
  \begin{tabular}{lcc}
    \toprule
    Method & Val, AUC & TestA, AUC\\
    \midrule
    \multicolumn{2}{l}{\emph{Video models} (See Section~\ref{sec:deepfake-classification} for Features)} & \\
    TCN with Features 1,2 & 84.29 & 69.31 \\
    TCN with Features 1,2,3 & 85.48 & 72.00 \\
    TCN with Features 1,2,3,4 & \textbf{88.41} & \textbf{73.11} \\
    \midrule
    \emph{Audio models} &  & \\
    Baseline Wav2Vec-AASIST-1 & 73.48 & 60.53 \\  
    Baseline Wav2Vec-AASIST-2 & 73.34 & --- \\  
    Wav2Vec-AASIST-RTPG & 99.35 & 82.14\\
    Wav2Vec-AASIST-BP & 99.46 & 82.57\\
    Wav2Vec-AASIST-codecs & \textbf{99.71} & \textbf{82.91} \\
    Wav2Vec-AASIST-all-aug & 98.20 & 80.58 \\
    Wav2Vec-AASIST-ensemble & 99.60 & 81.89 \\
    \midrule
    \emph{Multimodal fusion} &  & \\
    W2V-AASIST-codecs + Video, AVG & 97.86 & 91.97 \\
    KLASSify & \textbf{98.04} & \textbf{92.78} \\
  \bottomrule
\end{tabular}
\end{table}

\begin{table*}[!htb]
  \caption{Temporal localization performance on AV-Deepfake1M++ TestA~\cite{cai2025avdeepfake1mlargescaleaudiovisualdeepfake}}
  \label{tab:localization-results}
  \begin{tabular}{lcccccccccc}
    \toprule
    Method & TestA, IoU & AP@0.5 & AP@0.75 & AP@0.9 & AR@95 & AR@50 & AR@30 & AR@20 & AR@10 & AR@5 \\
    \midrule
    TCN with 4 features & 0.1139 & - & - & - & - & - & - & -& - & -\\ 
    KLASSify-BAM (audio only) & 0.3536 & 0.5117 & 0.4017 & 0.1701 &	0.0416 & 0.4259 & 0.4259 & 0.4259 & 0.4259 & 0.4258 \\
    \bottomrule
  \end{tabular}
\end{table*}

\subsection{Training}
\textbf{Video Classification and Localization}
To train the video classification and localization model, we selected a subset of videos shorter than 10.24 seconds or 256 frames from the train split. The model is trained to perform binary classification, distinguishing between real and fake videos. During training, we used various data augmentations to improve generalizability. These include feature shifting, temporal dropout, index swap, full channel dropout, and Gaussian noise. We use the AdamW \cite{loshchilov2019decoupledweightdecayregularization} optimizer with an initial learning rate of $1 \times 10^{-3}$ and a weight decay of $1 \times 10^{-2}$. The model is trained using \textit{binary cross-entropy} loss. During training, the input to the model is a 2 dimensional vector comprising eight handcrafted features along with their first-order differences. We perform channel-wise normalization for each feature. The model is trained for 100 epochs, and the learning rate is scheduled using cosine annealing \cite{loshchilov2017sgdrstochasticgradientdescent}. Early stopping is applied if there is no improvement in validation performance for more than 10 consecutive epochs. Both classification and localization share the same training pipeline. The localization training pipeline includes a tagging head that produces logits for each frame across four tags - \textit{O (Outside), B (Begin), I (Inside)}, and \textit{L (Last)}. We train this head using a \textit{weighted cross-entropy} loss so that the model learns to focus on learning the hard-sample, i.e. the fake segments. The exact weightage is [0.05, 0.30, 0.30, 0.35] for Outside, Begin, Inside, and Last, respectively. The model that performs best on the validation set is selected for testing.

During validation, we used a sliding window of 256 frames with a stride of 128 frames to generate fine-grained predictions. Empirically, it was found that score averaging across all windows yields the best performance on the TestA set.

\textbf{Audio Classification} 
Deepfake audio classification systems in Table \ref{tab:classif-result} are based on Wav2Vec-AASIST~\cite{Kukanov_2024,Tak2022Automatic} backbone. 
The audio systems are trained following the same optimization method, both the SSL front-end and the AASIST back-end are trained jointly. We use Adam \cite{kingma2017adammethodstochasticoptimization} optimizer with an initial learning rate of $1 \times 10^{-6}$. Negative log-likelihood loss is optimized with 2 class logSoftmax outputs for real and fake. Training stops using early stopping if there is no improvement in validation performance for more than 15 epochs. 

For training the deepfake audio classification on AV-\-Deep\-fake\-1M++~\cite{cai2025avdeepfake1mlargescaleaudiovisualdeepfake}, we extract audio stream from all videos using \emph{torchvision}\footnote{\url{https://docs.pytorch.org/vision/main/generated/torchvision.io.read_video.html}} library. Since the input context window to the Wav2Vec-AASIST is $\sim\!4$ seconds, we selected a subset of videos shorter than $6$ seconds duration for training, to be sure that at least a single fake segment is covered by the context window. At the \emph{training stage}, if the audio sample is longer than 4 seconds, the window of 4 seconds is randomly cut. If the audio is shorter than 4 seconds, it is wrapped around with a chunk from the beginning. At the \emph{evaluation stage}, the sliding window of 4 seconds with a stride of 1 second is fed to the trained model.

We have experimented with various data augmentations, see Table \ref{tab:classif-result}; these are from the \emph{audiomentations}\footnote{\url{https://iver56.github.io/audiomentations}} library \textbf{RTPG} - simulated room impulse responses~\cite{Ko2017rirs}, time stretch, pitch shift, Gaussian noise; \textbf{BP} - band pass filters: BandPassFilter, BandStopFilter, LowPassFilter, HighPassFilter, HighShelfFilter, LowShelfFilter, PeakingFilter. The \textbf{codecs} augmentations are applied using the \emph{torchaudio} library with the Sox\footnote{\url{https://docs.pytorch.org/audio/0.11.0/_modules/torchaudio/backend/sox_io_backend.html}} back-end with the \emph{format} transformations of htk, gsm, amr-nb, sph, vorbis, ogg, mp3, amb, wav; \emph{bits per sample quantizations} 8, 16, 32; \emph{encodings}  PCM\_S, PCM\_U, PCM\_F, ULAW, ALAW. With the probability of $p=0.3$, we randomly sample a specific subset of these configurations and apply the transformation to an input audio file at the training stage.  
The system trained with all of the above augmentations is \textbf{all-aug} and the shallow \textbf{ensemble} (score-based averaged) of these systems is presented in Table \ref{tab:classif-result}.

\begin{figure}[!htb]
  \centering
  \includegraphics[scale=0.25]{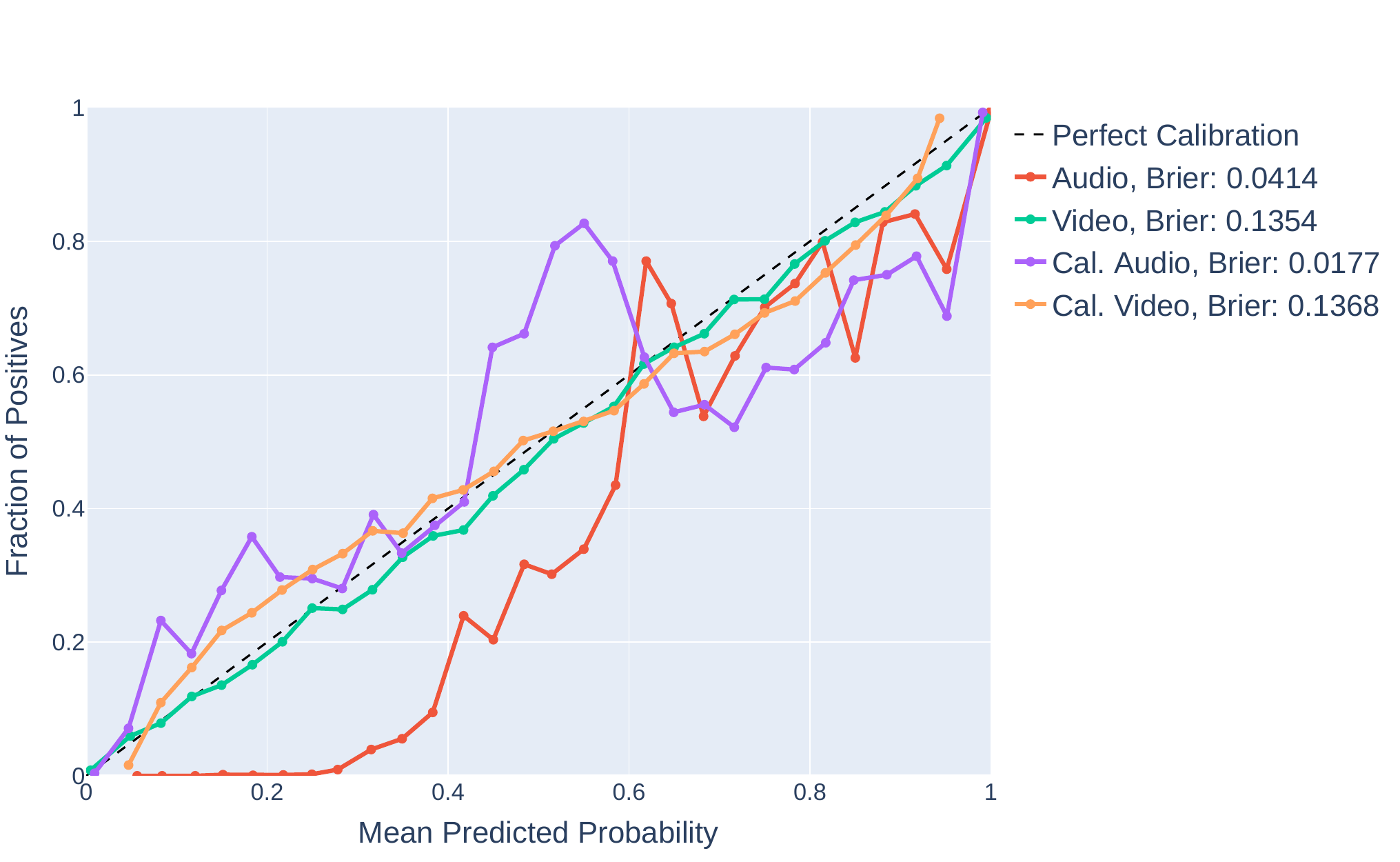}
  \caption{Calibration curves for audio and video models before/after Platt~\cite{Platt1999Probabilistic} score calibration }
  \label{img:cal_curves}
\end{figure}

\textbf{Classification Fusion} 
The Platt \cite{Platt1999Probabilistic} calibration is trained on the 40\% of the validation split of AV-Deepfake1M++ \cite{cai2025avdeepfake1mlargescaleaudiovisualdeepfake} dataset. Comparison of un-calibrated and calibrated scores are in Figure \ref{img:cal_curves}.

\textbf{Audio Localization} 
For the fake audio localization, we train the Boundary-aware Attention Mechanism (BAM) \cite{zhong2024enhancingpartiallyspoofedaudio} model on AV-Deepfake1M++ \cite{cai2025avdeepfake1mlargescaleaudiovisualdeepfake} dataset. Following the configuration in~\cite{zhong2024enhancingpartiallyspoofedaudio}, the model employs a pre-trained self-supervised front-end WavLM-Large~\cite{journals/corr/abs-2110-13900} as the feature extractor, followed by an attentive pooling layer with a $40$ms temporal frame resolution. The pooled features are then processed by the \emph{boundary enhancement} (BE) and the \emph{boundary frame-wise attention} modules. The overall model is trained by optimizing a combined loss function of \emph{boundary loss} and \emph{frame-level authenticity loss} with the weighting parameter of $0.5$ \cite{zhong2024enhancingpartiallyspoofedaudio}. 

The maximal duration of audio files selected for training the BAM model is 6 seconds. The model input is a random cut window of 4 seconds, which corresponds to the maximal label length of $100$ frames with the resolution of $40$ms per frame ($4 / 0.04 = 100$). We employ the Adam~\cite{kingma2017adammethodstochasticoptimization} optimizer with an initial learning rate of $1 \times 10^{-5}$ that is then halved every $10$ epochs. We also employ early stopping if the validation loss fails to reduce for $3$ epochs. The full BAM model is pre-initialized with the checkpoint weights from \cite{zhong2024enhancingpartiallyspoofedaudio} trained on PartialSpoof~\cite{zhang21ca_interspeech} dataset. 

\section{Results and Analysis}
\textbf{Video Classification}
For the video classification, we experimented with four different feature groups. As shown in Table \ref{tab:classif-result}, training a TCN~\cite{lea2016temporalconvolutionalnetworksunified} based on \textit{Blurriness of the mouth ROI} and \textit{Non-mouth MSE} achieved AUC of $84.29\%$ and $69.31\%$ on the validation and testA sets, respectively. Incorporating additional handcrafted features, such as \textit{Color Shift} and \textit{Landmark Kinematics}, improved the AUC to $73.11\%$ on testA, which is roughly $18\%$ gain over the baseline Xception~\cite{Chollet_2017_CVPR} model. These experimental results support the effectiveness of handcrafted features for video deepfake detection.

\textbf{Audio Classification}
In Table \ref{tab:classif-result}, initial baseline models, \textit{Baseline Wav2Vec-AASIST-1}~\cite{Kukanov_2024} and \textit{Baseline Wav2Vec-AASIST-2} are trained on ASVspoof 2019 (LA) \cite{Todisco2019ASVspoof2F} and ASVspoof 2024 \cite{Wang2024ASVspoof5C} datasets, respectively, under the scenario when an utterance is entirely real or fake. \textit{Conventional utterance-level approaches, designed for the fully-fake-utterances, largely fail in the face of partially-spoofed utterances, indicating their vulnerability to such attacks.} Both baselines achieve AUC of $\sim\!73\%$ and $63.53\%$ on the validation and testA sets. This finding confirms the results that the authors concluded in \cite{zhang23_partialspoof_taslp} that segment-level detection is a more challenging task than utterance-level detection due to shorter fake segments. 

Further training on the AV-Deepfake1M++ \cite{cai2025avdeepfake1mlargescaleaudiovisualdeepfake} dataset boosts the model's AUC performance to $99.71\%$ and $82.91\%$ on the validation and testA sets, making it the best-performing single Wav2Vec-AASIST model with codecs augmentations. It slightly outperforms both the model trained with all augmentations (Wav2Vec-AASIST-all-aug) and even the ensemble of audio models (Wav2Vec-AASIST-ensemble).

\textbf{Multimodal Calibration \& Fusion}
In Table \ref{tab:classif-result}, the video and audio models, best performing on the validation set, were selected for the calibration, the multimodal fusion, and the final submission. In Figure~\ref{img:cal_curves}, the calibration curves are presented before and after Platt~\cite{Platt1999Probabilistic} score scaling. We noticed that video scores are close to perfect calibration, while audio scores improved calibration after Platt~\cite{Platt1999Probabilistic} scaling. 

\textit{KLASSify} solution for deepfake classification comprises audio Wav2Vec-AASIST-codecs and video TCN model, followed by calibration and max-out final score decision. It outperforms regular audio-video model score averaging (\textit{W2V-AASIST-codecs + Video, AVG}) from AUC of $91.97\%$ to $92.78\%$ on testA split.

\textbf{Video Localization}
Initial experiments showed promising results, in Table \ref{tab:localization-results}, achieving the IoU of $0.1139$ on testA, using only handcrafted video features. Our approach was $3\%$ lower than the baseline method, BA-TFD+ \cite{BATFD+} results the IoU of $0.1471$, which utilized heavyweight transformers and multimodal features. Further exploration is needed to improve the video localization task.

\textbf{Audio Localization}
Prior to training on AV-Deepfake1M++ \cite{cai2025avdeepfake1mlargescaleaudiovisualdeepfake} dataset, we explored the generalization of the BAM~\cite{zhong2024enhancingpartiallyspoofedaudio} model trained on PartialSpoof~\cite{zhang21ca_interspeech} dataset. Our experiments revealed that the model struggled to generalize to fake segments localization on AV-Deepfake1M++~\cite{cai2025avdeepfake1mlargescaleaudiovisualdeepfake}. This finding is consistent with prior research in \cite{zhang2025PartialEdit}, which reported that deepfake localization systems underperform when trained and evaluated on mismatch acoustic environment or unseen deepfake attacks. The BAM model checkpoint from \cite{zhong2024enhancingpartiallyspoofedaudio} achieved EER of only $77.52\%$ of fake frames detection on AV-Deepfake1M++ validation split. Further finetuning on AV-Deepfake1M++ \cite{cai2025avdeepfake1mlargescaleaudiovisualdeepfake} improved the EER to $3.20\%$ or IoU of $0.6750$ on the validation split. The final scores of the \textit{KLASSify-BAM} model in Table \ref{tab:localization-results} were submitted to the leaderboard and achieved the IoU of $0.3536$ on testA set.

\section{Conclusion}
In this work, we presented KLASSify, a computationally efficient and robust system for temporal deepfake detection and localization on the AV-Deepfake1M++~\cite{cai2025avdeepfake1mlargescaleaudiovisualdeepfake} dataset. Despite its simplicity, the proposed multimodal audiovisual fusion approach achieves competitive performance, attaining the AUC of $92.78\%$ on the deepfake classification task on testA data split. Furthermore, we explored a boundary-aware attention approach \cite{zhong2024enhancingpartiallyspoofedaudio} for the localization task, achieving the IoU of $0.3536$ using only the audio modality.

Despite the competitive results, several open challenges remain in the field. First, utterance-based deepfake detection systems do not generalize well to partial deepfake classification. The partial fake segments are short and making task more challenging than full-utterance deepfake detection. Second, cross-dataset generalization of deepfake classification and localization remains an issue. In our experiments, a deepfake localization system pre-trained on PartialSpoof~\cite{zhang21ca_interspeech} failed to generalize on AV-Deepfake++~\cite{cai2025avdeepfake1mlargescaleaudiovisualdeepfake} dataset without additional finetuning. Both of these problems were confirmed experimentally and require further investigation in future work.

\newpage

\clearpage


\printbibliography

\end{document}